\documentclass[a4paper,aip,pof,amsmath,amssymb,
 reprint,%
]{revtex4-1-per}
\LogoHeight{1.85cm}

\usepackage{graphicx}
\usepackage{dcolumn}
\usepackage{bm}
\usepackage{pifont} 
\usepackage[utf8x]{inputenc}
\usepackage{xcolor}

\begin{document}

\preprint{AIP/123-QED}

\title{Turbulent velocity spectra in superfluid flows}


\author{J. Salort$^1$}

\author{C. Baudet$^2$}
\author{B. Castaing$^3$}
\author{B. Chabaud$^1$}
\author{F. Daviaud$^4$}
\author{T. Didelot$^1$}
\author{P. Diribarne$^{1,3,4,5}$}
\author{B. Dubrulle$^4$}
\author{Y. Gagne$^2$}
\author{F. Gauthier$^1$}
\author{A. Girard$^5$}
\author{B. Hébral$^1$}
\author{B. Rousset$^5$}
\author{P. Thibault$^{1,5}$}

\author{P.-E. Roche$^1$}

\affiliation{$^{1)}$Institut Néel, CNRS/UJF - 25 rue des Martyrs, BP 166, F-38042 Grenoble cedex 9\\
$^{2)}$Laboratoire des Écoulements Géophysiques et Industriels, CNRS/UJF/INPG, F-38041 Grenoble cedex 9\\
$^{3)}$École Normale Supérieure de Lyon - 46 allée d'Italie, F-69364 Lyon cedex 7\\
$^{4)}$DSM/IRAMIS/SPEC, CEA Saclay and CNRS (URA 2464), F-91191 Gif sur Yvette cedex \\
$^{5)}$CEA-Grenoble/UJF, SBT, 17 rue des Martyrs, BP166, F-38054 Grenoble cedex 9}

\date{\today}

\begin{abstract}

  We present velocity spectra measured in three cryogenic liquid
  $^4\mathrm{He}$ steady flows: grid and wake flows in a pressurized wind
  tunnel capable of achieving mean velocities up to 5~m/s at
  temperatures above and below the superfluid transition, down to
  1.7~K, and a ``chunk'' turbulence flow at 1.55~K,
  capable of sustaining mean superfluid velocities up to 1.3~m/s.
  Depending on the flows, the stagnation pressure probes used for
  anemometry are resolving from one to two decades of the inertial regime
  of the turbulent cascade. We do not find any evidence that the
  second order statistics of turbulence below the superfluid transition
  differ from the ones of classical turbulence, above the transition.

\end{abstract}



\maketitle

\section{Introduction}

At atmospheric pressure and below approximately $4.2~\mathrm{K}$,
$^4\mathrm{He}$ forms a liquid phase, called He~I, whose dynamics can be
described by the Navier-Stokes equation.  When this liquid is cooled
below $T_{\lambda} \approx 2.17~\mathrm{K}$, it undergoes a phase
transition, the ``superfluid'' transition. The new liquid
phase is called He~II. The hydrodynamics of this phase can be described
with the so-called two-fluid model\cite{landauVol6}, ie. as a
superposition of a normal component which behaves like a classical
Navier-Stokes fluid with finite viscosity and a superfluid one with
zero-viscosity and quantized vorticity.  The ratio of superfluid
density versus total density, $\rho_s/\rho$ increases from 0 to 1 when
temperature decreases from $T_{\lambda}$ to 0~K (typical values are
given in table \ref{tab:HeProp}).  The main goal of this paper is to
compare the statistics of turbulent flows above and below this
``superfluid'' transition.

To achieve this goal, we need a local sensor that can work both above
and below $T_{\lambda}$. Unfortunately, the most efficient sensors
available, can only operate in one of these phases, hot-wires for
$T>T_{\lambda}$\cite{castaing1994_turbulence,zocchi1994,chanal2000,pietropinto2003},
and quantum vortex lines density probes for
$T<T_{\lambda}$\cite{holmes1992,smith1993,stalp2002,skrbek2003,roche2007}.



One alternative possibility is to use stagnation pressure probes.  The
operating principle is similar to Pitot or Prandtl tubes:  the
velocity difference between the tip of the probe where the flow is
stopped and the average flow velocity produces a pressure head
$\frac{1}{2}\rho v^2$. This effect is inertial, and therefore such
probes can be used as well in He~I as in He~II.


The first successful attempt to resolve velocity fluctuation in liquid
helium with a stagnation pressure probe was reported in 1998 by Maurer
and Tabeling\cite{maurer1998} in a turbulent Von Kármán flow both
above and below $T_{\lambda}$.  The velocity spectra in He~II were
found very similar to those in He~I. Specifically they found a
$f^{-5/3}$ scaling over 1.5 decade of frequency.  This pioneering
result provides the first experimental evidence that superfluid can
undergo a Kolmogorov-like turbulent cascade.  Yet, there has been no
published experimental confirmation of this result\footnote{The
  confirmation previously cited by Roche, \emph{et al.}\cite{roche2007} is
  presented in the present paper}. For reference, we point that
numerical works have reported spectrum compatible with a -5/3 scaling
at finite temperature\cite{merahi2006,roche2009} and in the zero temperature
limit\cite{nore1997,araki2002,kobayashi2005}. The reader can
report to the review of Vinen and Niemela for an introduction to
quantum turbulence\cite{vinen2002}.

This paper presents an extension of this experimental result in
different geometries. We report studies of stagnation pressure
measurements both in He~I and He~II for three kinds of flow: grid
turbulence, wake near field flow and ``chunk'' flow
with two objectives in mind: (i) to check that the
experiment when done in a classical fluid like He~I reproduces
expected statistical signatures for the turbulence and (ii) to
compare the statistical signatures for flows in He~I with those
in He~II.

\begin{table}
\caption{\label{tab:HeProp}Some physical properties of cryogenic helium for temperature
and pressure values relevant to our experiments}
\begin{ruledtabular}
\begin{tabular}{ccccc}
$P$ [Pa] & $T$ [K] & $\rho$ [$\mathrm{kg/m^3}$] & $\eta$ [$\mathrm{\mu Pa\cdot s}$] & $\rho_s/\rho$ \\
\hline\hline
\multicolumn{5}{c}{\emph{Pressurized He~I}} \\
$1.1\times 10^5$ & 2.6 & 146.6 & 3.374 & 0 \\
$1.1\times 10^5$ & 2.3 & 148.0 & 2.980 & 0 \\
\hline
\multicolumn{5}{c}{\emph{Pressurized He~II}} \\
$1.1\times 10^5$ & 2.17 & 148.2 & 2.611 & 0 \\
$1.1\times 10^5$ & 2.1 & 147.7 & 1.971 & 0.23 \\
$1.1\times 10^5$ & 2.0 & 147.5 & 1.555 & 0.42 \\
$1.1\times 10^5$ & 1.9 & 147.3 & 1.389 & 0.56 \\
$1.1\times 10^5$ & 1.7 & 147.1 & 1.359 & 0.76 \\
\hline
\multicolumn{5}{c}{\emph{Saturated He~II}} \\
597 & 1.55 & 145.3 & 1.380 & 0.86
\end{tabular}
\end{ruledtabular}
\end{table}

\section{Probes and acquisition system}

In this paper, we report measurements done with four stagnation
pressure probes, hereafter called \ding{172}, \ding{173}, \ding{174} and
\ding{175}. They were used in two wind tunnels (described below), noted
TSF and NÉEL for convenience.  Two types of
pressure transducers were used, Kulite cryogenic ultraminiature
CCQ-062 pressure transducers for probes \ding{172} and \ding{174}, and
a Fujikura Ltd. FPS-51F-15PA pressure
transducer\cite{haruyama1998,maeda2004} for probes \ding{173} and
\ding{175}.  Both transducers are based on piezoresistive gauges.

\begin{figure}
\hspace*{\stretch{1}}\begin{minipage}{2cm}
\centering\includegraphics[width=1cm]{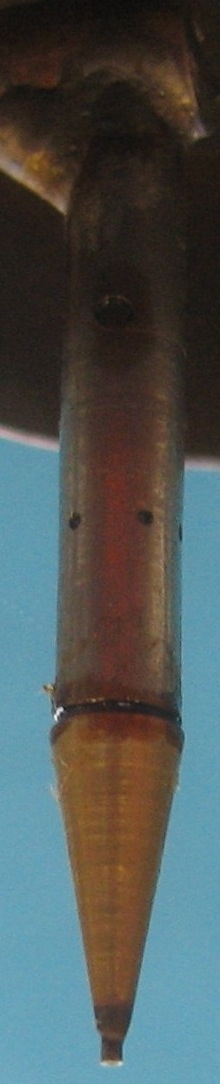}
\end{minipage}\begin{minipage}{2.3cm}
\includegraphics[scale=0.8]{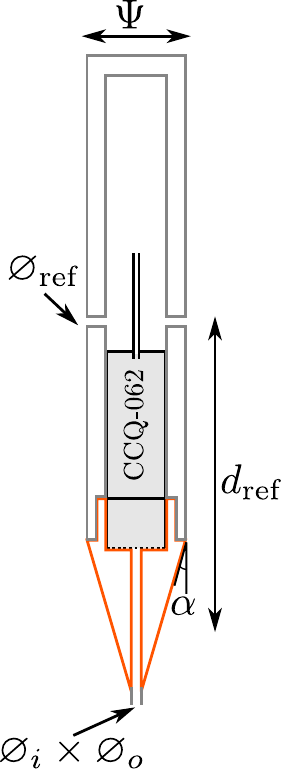}
\end{minipage}\hfill\begin{minipage}{4.1cm}
\centering\includegraphics[width=2cm]{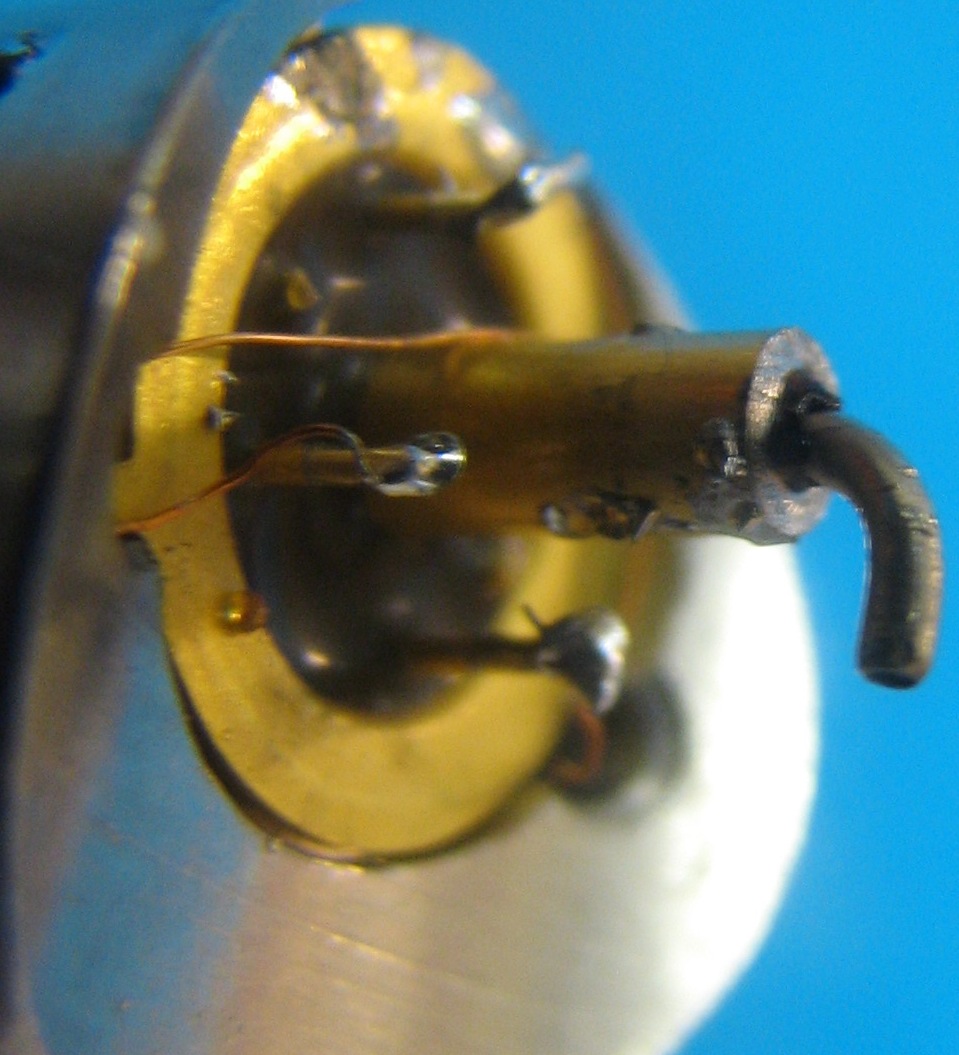}
\vspace{0.5cm}
\includegraphics[scale=0.8]{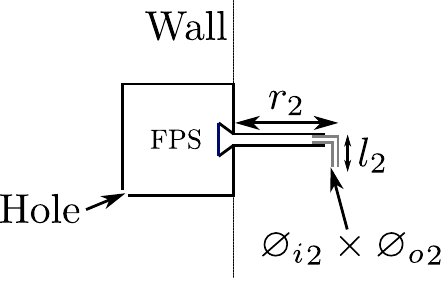}
\end{minipage}\hspace*{\stretch{1}}

\vspace{0.2cm}

\hspace*{\stretch{1}}\begin{minipage}{4.3cm}
\centering (a)
\end{minipage}
\hfill\begin{minipage}{4.1cm}
\centering (b)
\end{minipage}

\vspace{0.5cm}

\hspace*{\stretch{1}}
\includegraphics[height=2.5cm]{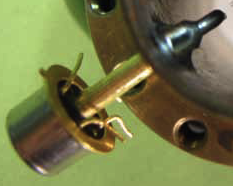}\hfill
\includegraphics[height=2.5cm]{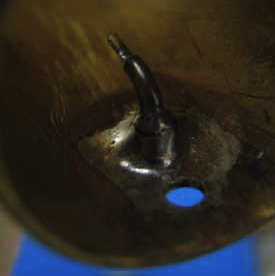}
\hspace*{\stretch{1}}

\vspace{0.5cm}

\centering (c)

\caption{\emph{TSF wind tunnel probes}: (a) probes \ding{172} or \ding{174};
  (b) probe \ding{173}. All parts are tightly assembled. For probe
  \ding{172} and \ding{174}, the pressure reference is realised by
  holes on the outer CuNi cylinder at a distance $d_{\text{ref}}$ from
  the tip; for probe \ding{173}, the pressure reference is taken in a
  region where the flow is quiescent with a controlled leak on the
  back of the shell.  \emph{NÉEL wind tunnel probe}: (c) Probe
  \ding{175} is essentially similar to probe \ding{173} except that it works as
  an absolute pressure probe, without hole in its shell.}
\label{fig:probeSupports}
\end{figure}

They have been customized by wrapping them into specifically designed
noses and supports in order to get a smaller resolution. The tips of the
noses are made of cupro-nickel capillaries, of typical diameter
$0.4\times 0.6~\mathrm{mm}$ for probes \ding{172} and \ding{174}, and
$0.6\times 0.9~\mathrm{mm}$ for probe \ding{173} and \ding{175} (see
figure \ref{fig:probeSupports}). The nozzle sizing is optimized for
space and time resolution.  In first approximation, the nozzle acts as
a pipe and the dead volume inside the Kulite CCQ-062 outfit as a
cavity. This introduces a Helmholtz resonance for probes \ding{172} and
\ding{174}.  For probes \ding{173} and \ding{175}, the dead volume is
negligible but the pipe total length is typically 1~cm, leading to an
organ pipe resonance.  For probes \ding{172}, \ding{173} and
\ding{174}, the resonance frequency is found around 2~kHz, which means that,
for a mean flow velocity of 1~m/s, we cannot resolve structures smaller
than 1~mm typically. 
For probe \ding{175}, the resonance frequency is below 1~kHz.
The time and space cut-off of all the probes therefore occurs simultaneously.

\begin{figure}
\centering \includegraphics[scale=0.75]{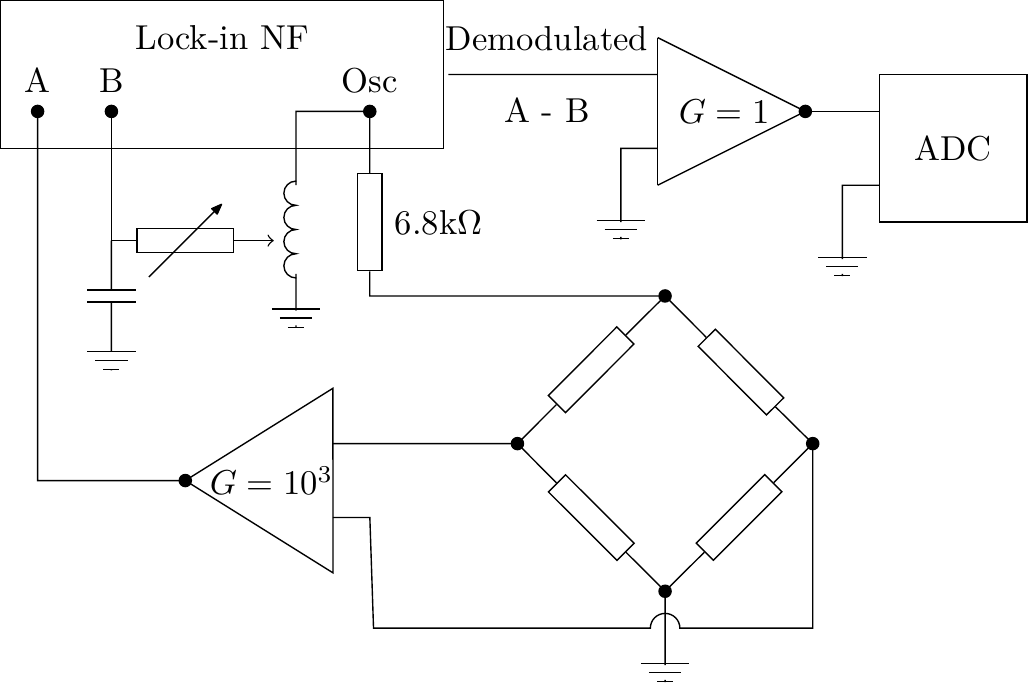}

\caption{Electronic diagram of the acquisition system for probes
  \ding{172} and \ding{173}. The pressure transducer is represented as
  a resistor bridge. The imbalance is preamplified by a low-noise
  preamplifier (JFET, typical noise input voltage
  $1~\mathrm{nV/\sqrt{Hz}}$).  The mean value of the imbalance signal
  is substracted using an inductor bridge and an
  adjustable RC filter to compensate for the phase shift. 
}\label{fig:electronique}
\end{figure}

Probes \ding{172}, \ding{173} and \ding{175} have been polarized with
a sinusoidal voltage. The output signal is demodulated by
a 
lock-in amplifier. The polarisation frequency is in the range 7 ---
8~kHz for probes \ding{172} and \ding{173} and in the range 10 ---
20~kHz for probe \ding{175}.  This modulation/demodulation technique
was chosen to improve the signal to noise ratio. To make sure that no
artefact bias was introduced by this method, probe \ding{174} was
polarized more simply using DC batteries. The full acquisition
schematics is given on figure \ref{fig:electronique}. The various properties
of the probes are summarized in table \ref{tab:Probes}.

\begin{table}
\caption{\label{tab:Probes}Summary of the main properties of the probes used in
our experiments}
\begin{ruledtabular}
\begin{tabular}{rcccc}
Probe & \ding{172} & \ding{173} & \ding{174} & \ding{175} \\
\hline\hline
Transducer & Kulite & Fujikura & Kulite & Fujikura \\
Nose diameter [mm] & $0.4\times 0.6$ & $0.6\times 0.9$ & $0.4\times 0.6$ & $0.6\times 0.9$\\
Resonance [kHz] & $\sim 2$ & $\sim 2$ & $\sim 2$ & $< 1$ \\
Sensing & AC & AC & DC & AC
\end{tabular}
\end{ruledtabular}
\end{table}

\section{Stagnation pressure probes used as anemometers}

Following the analysis of Maurer and Tabeling\cite{maurer1998}, the
first order term of the signal fluctuations measured by a stagnation
pressure probe is linear with the local velocity fluctuations, like
with Pitot tubes. However, if the turbulence intensity is too large,
the second-order corrections coming from static pressure fluctuations and
quadratic velocity fluctuations lead to significant bias (see appendix
\ref{app:Calculs} for more details).

Maurer and Tabeling's measurements were done using a stagnation
pressure probe inside a turbulent Von Kármán flow.  The piezoelectric
probe they used was not sensitive to the DC but they could measure the
turbulence intensity $\tau$ in the range 20 --- 30~\% in a previous
measurement\cite{maurer1994}. According to table \ref{tab:terms}, in
such conditions, the second-order corrections represent more than
$\sim 20~\%$ of the measured signal.
Additionnally, events with flow-probe angle of attack exceeding for
example 15° are likely to occur at such high $\tau$, which introduces
some additionnal bias on the signal interpretation.  To confirm and
extend Maurer and Tabeling's result, our systematic study includes a
flow with a turbulence intensity smaller than 2~\%, with second-order
correction smaller than 3~\%. A grid
flow was chosen because its turbulence is well known in classical
fluids.

The calibration of the probes is done \emph{in-situ}, by plotting the
mean output voltage versus
$\rho\left< v \right>^2$ where $\left< v\right>$ is the mean velocity
in the channel.  In the TSF wind tunnel, $\left<v\right>$ is
determined by enthalpy balance across a heater.  In the NÉEL wind tunnel, a
Pitot tube located downstream from the probe (see figure
\ref{fig:schemaTSF}) provides a measurement of the flow mean velocity.

\section{Homogeneous and isotropic turbulence: the TSF grid flow}

\begin{figure*}
  \hspace*{\stretch{1}}
  \begin{minipage}{.51\textwidth}
    \begin{minipage}{.5\textwidth}\centering
      \includegraphics[scale=0.6]{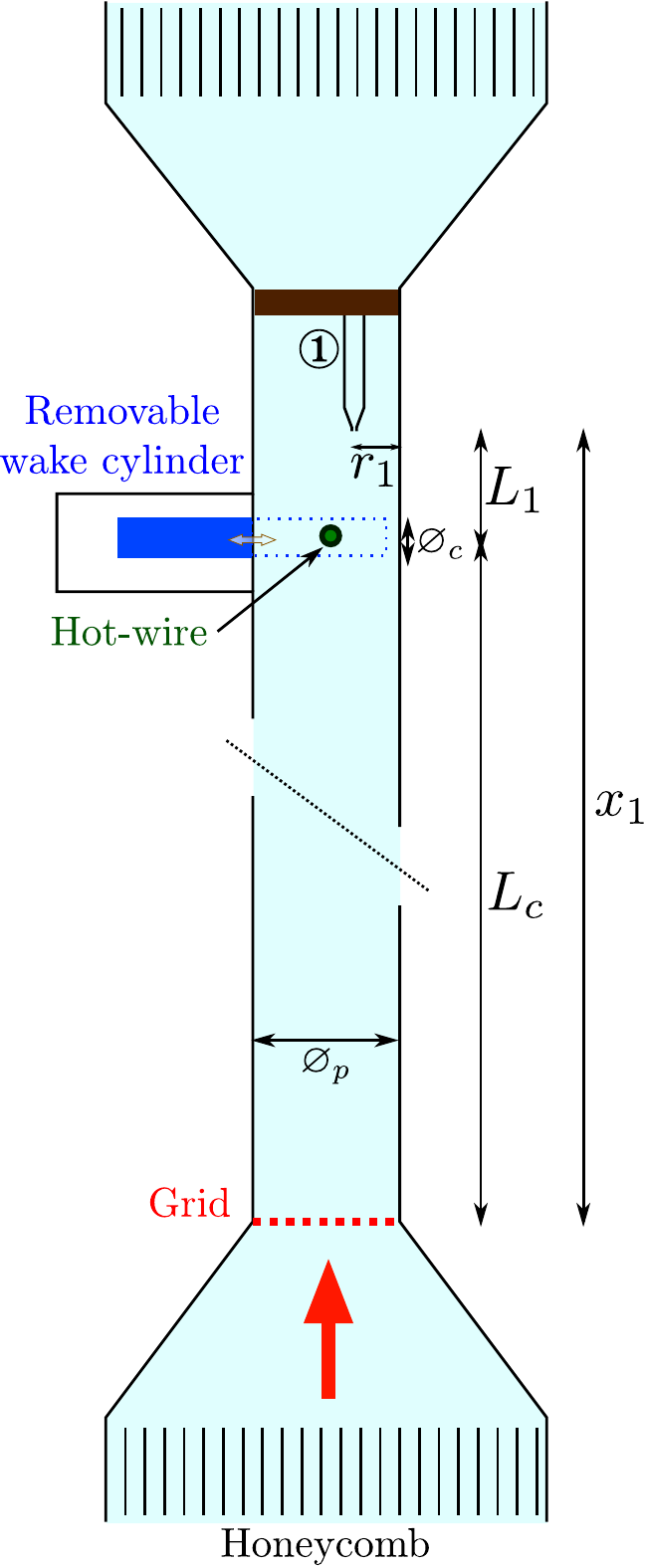}\\
      Run 1
    \end{minipage}\hfill
    \begin{minipage}{.5\textwidth}\centering
      \includegraphics[scale=0.6]{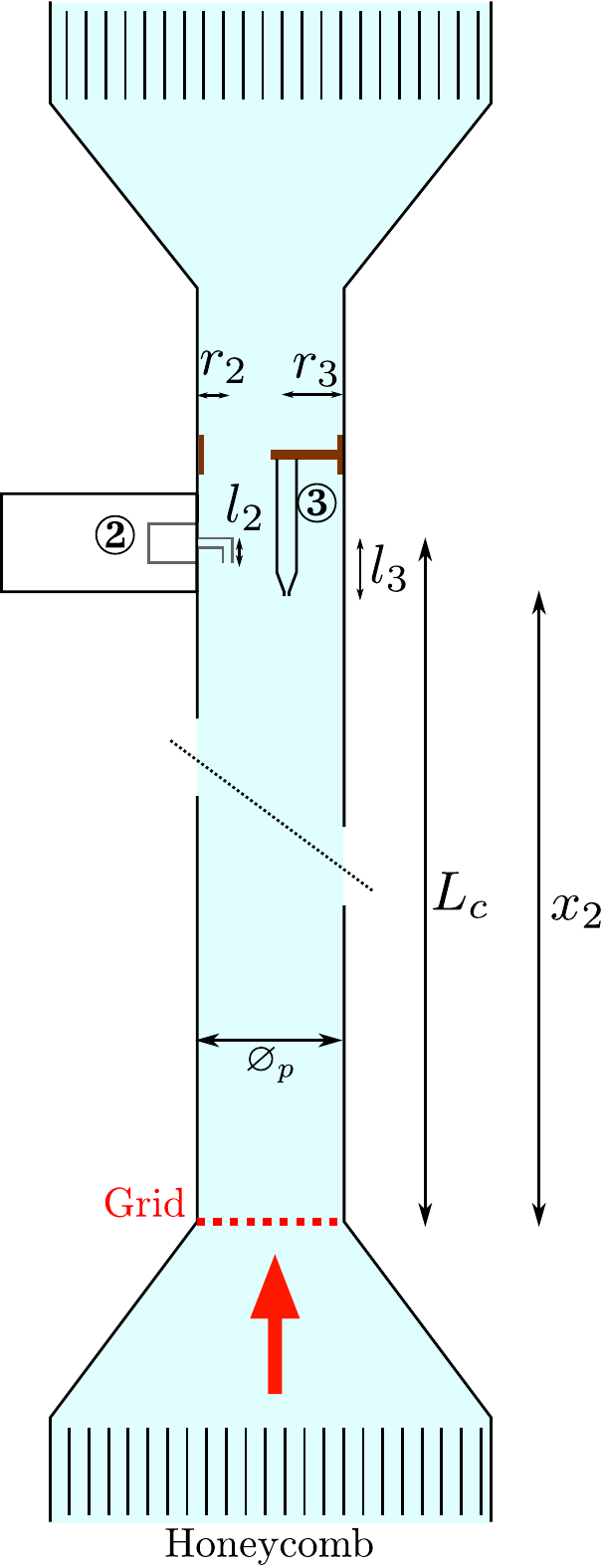}\\
      Run 2
    \end{minipage}\\
    \centering (a)
  \end{minipage}\hfill
  \begin{minipage}{.47\textwidth}
    \centering
    \includegraphics[width=\textwidth]{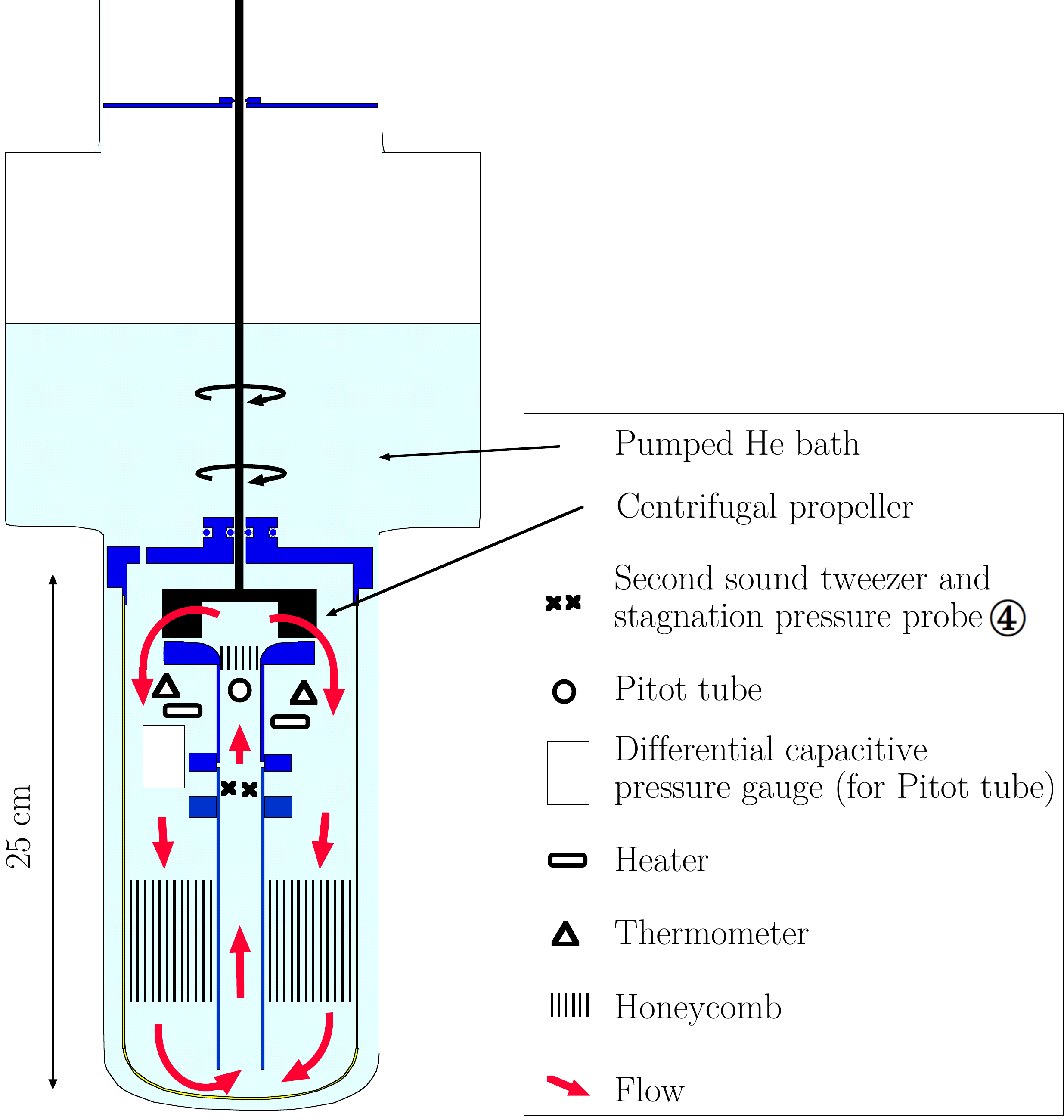}\\
    (b)
  \end{minipage}\hspace*{\stretch{1}}

  \caption{\emph{(a) TSF wind tunnel}: Schematics of the test section and the
  probe locations for runs 1 and 2. For run 1, a removable cylinder
  can be inserted across the flow at a distance $L_c$ downstream the grid.
  It was originally designed to protect a hot-wire during the
  transient of the system.  The stagnation pressure probe \ding{172},
  located at a distance $L_c+L_1$ downstream the grid can either measure
  grid turbulence when the cylinder is removed or wake turbulence when
  the cylinder is inserted in the flow. Probe \ding{172} was not
  positionned on the pipe axis to avoid the wake of the hot-wire. For
  run 2, two stagnation pressure probes (\ding{173} and \ding{174})
  are available. 
  \emph{(b) NÉEL wind tunnel}: Schematics and picture of the test section and
  location of stagnation pressure probe \ding{175}.}
\label{fig:schemaTSF}
\end{figure*}

In this section, we present grid turbulence measurements in the
pressurized TSF wind tunnel (see figure \ref{fig:schemaTSF}).  Details
about the TSF experiment have been given in previous
papers\cite{rousset2008, diribarne2009}. The main dimensions are
recalled in table \ref{table:dimTSF}.  The turbulence intensity in
this type of flow is typically a decade smaller than turbulent Von
Kármán flows, which ensures that the fluctuating signal from the
stagnation pressure probes corresponds to velocity fluctuations with
less than 3~\% correction. Furthermore, the pressure is maintained far
above the satured vapor pressure, this ensures that no bubble can
appear within the flow.  However, one drawback of low turbulence
intensity is that the fluctuating signal on the probe is lower, therefore the
signal to noise ratio is smaller.  

In this paper, we discuss two runs with different probe
positions inside the test section (shown on figure
\ref{fig:schemaTSF}), with mean velocities ranging from 0.4~m/s to
5~m/s and temperatures from 1.7~K and 2.6~K. The Reynolds number based
on the grid mesh size $M$, $\mathrm{Re}_M = M\left<v\right>/\nu$ is
between $10^5$ and $2\cdot 10^6$ in He~I. In He~II, several Reynolds
numbers can be defined. Using the quantum of circulation $\kappa =
h/m \simeq 9.9\times 10^{-8}~\mathrm{m^2/s}$ ($h$ is the Planck
constant and $m$ is the mass of the $^4\mathrm{He}$ atom), we find
$\mathrm{Re}_{\kappa} = M\left<v\right>/\kappa$ between $1.5\times
10^4$ and $2\times 10^5$.

The probe location downstream the grid is $x/M = 138 \pm 2$ for the first
run and $x/M = 121 \pm 2$ for the second run. Hence we can derive the
turbulence intensity and the transverse integral scale $L_g$ expected
in He~I using Comte-Bellot and Corrsin's fits \cite{comteBellot1966},
\begin{equation}
\left<v\right>^2/\left< v'^2 \right> = 15\left( \frac{x}{M} - \frac{x_0}{M} \right)^{1.2}
\label{eq:comteBellot:TurbIntensity}
\end{equation}
\begin{equation}
L_g/M = 0.06\left( x/M - x_0/M \right)^{0.35}
\label{eq:comteBellot:IntegralScale}
\end{equation}
where $x_0/M$ is the virtual origin ranging from 2 to 4.

The expected turbulence intensity in the TSF loop is therefore between
1.3~\% and 1.5~\% and the expected transverse integral scale $L_g$
lies in the range 1.2 --- 1.3~mm.  Alternative prefactors and
exponents in equations \ref{eq:comteBellot:TurbIntensity} and
\ref{eq:comteBellot:IntegralScale} have been proposed in the
literature. Using those reported by Mohamed and LaRue\cite{mohamed1990}, we find
a turbulence intensity between 0.92~\% and 1.7~\% for $x/M = 121$ (run
2) and 0.84~\% and 1.6~\% for $x/M=138$ (run 1). In any cases, the turbulence
intensity is small enough to safely assume that the measured signal is
not polluted by static pressure fluctuations nor by large angle of attack between
the flow and the probe.

\begin{table}
  \caption{\label{table:dimTSF}Main dimensions of the TSF wind tunnel (see figure \ref{fig:schemaTSF} and \ref{fig:probeSupports} for the definition of the notations)}
\begin{ruledtabular}
\begin{tabular}{rlrlrlrl}
$\varnothing_p$ & 27.2~mm & $L_1$ & 61~mm  & $l_2$ & 3~mm & $l_3$ & 9~mm \\
$\varnothing_c$ & 15.3~mm & $r_1$ & 8~mm  & $r_2$ & 7~mm & $r_3$ & 11~mm\\
$L$ & 565~mm & ${\varnothing_i}_1$ & 0.4~mm  & ${\varnothing_i}_2$ & 0.6~mm & ${\varnothing_i}_3$ & 0.4~mm \\
$L_c$ & 479~mm & ${\varnothing_o}_1$ & 0.6~mm & ${\varnothing_o}_2$ & 0.9~mm & ${\varnothing_o}_3$ & 0.6~mm \\
$M$ & 3.9~mm/mesh & $n_M$ & 7~mesh/diam & $\varnothing_{\text{ref}}$ & 0.5~mm & $d_{\text{ref}}$ & 15~mm \\
$\Psi$ & 3.5~mm & $\alpha$ & 15° & & & & \\
\end{tabular}
\end{ruledtabular}
\end{table}

\begin{figure}
\begin{center}
\includegraphics{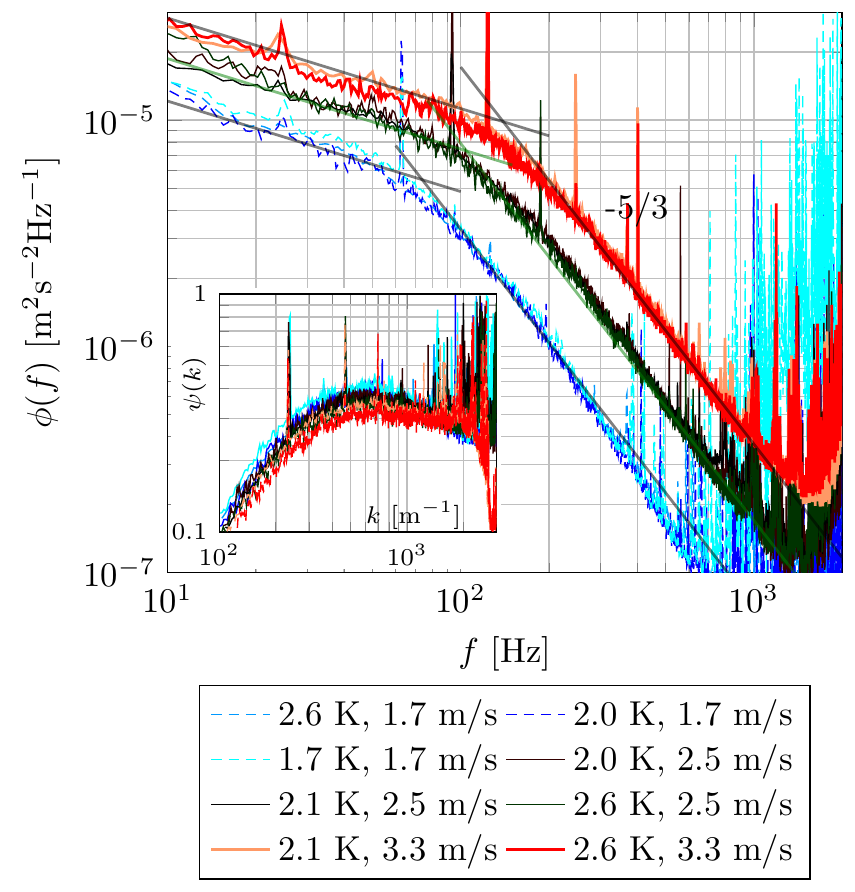}
\end{center}
\caption{Grid turbulence velocity spectra acquired by probe \ding{172}
  for three different mean velocities both above and below the
  superfluid transition. The Helmholtz resonance frequency is found
  near 2~kHz. The solid lines are visual aids to find the corner
  frequency, $f_0$.  The high-frequency lines show the -5/3 scaling.
  Inset: Compensated grid flow energy spectrum for various conditions
  both above and below the superfluid transition (see text). The value
  of the plateau provides an estimate for the one-dimensional
  Kolmogorov constant $C_k$ for both He~I and He~II grid turbulence.}
\label{fig:spectra}
\end{figure}

The velocity power spectra, $\phi(f)$, are given on figure \ref{fig:spectra}, where
the velocity spectral density over the time interval $[ 0, t_{\text{max}} ]$, $\phi(f)$, is defined as
\begin{equation}
\phi(f) = \left| \sqrt{\frac{2}{t_{\text{max}}}}\int_{0}^{t_{\text{max}}}v'(t)e^{-2i\pi ft}\mathrm{d}t \right|^2
\end{equation}
The normalisation is such that
\begin{equation}
\int_0^{+\infty}\phi(f)\mathrm{d}f = \left< v'^2 \right>
\end{equation}
The actual spectra are calculated using the Welch method on windows of
$2^{15}$ data points. The total recording time is 7~min for most time
series but we also recorded some 15~min and 30~min-long ones,
with a sampling frequency of 9.77~kHz or 19.5~kHz and a high-order
antialiasing filter.  In He~I, a
Kolmogorov scaling $\phi(f) \sim f^{-5/3}$ is expected in the inertial
range of the turbulent cascade. Above the corner frequency around 100
--- 200~Hz, our measurements are compatible with such a scaling
although the limited resolved range calls for caution.
On this representation, the measurements in He~II seem
indistinguishable from those in He~I, which suggests that the
turbulence second-order statistics in the upper part of the inertial
cascade are the same above and below the superfluid
transition. However, this representation is not well suited for
detailed comparisons because of the peaks of noise. In the following,
we present more quantitative characteristics of this spectra to refine
the comparison of  flows in He~I and He~II, ie below and above the superfluid
transition.

We first examine the integral scale of the flow and
the turbulence intensity. Both can be calculated from
the spectra. The values obtained above the superfluid
transition can be compared against Comte-Bellot and Corrsin's
fits for classical grid flows.

The longitudinal integral scale
in the flow, $L_l$, can be defined as
\begin{equation}
L_l = \frac{1}{\left<v'^2\right>}\int_0^{+\infty}\left<v'(0)v'(r)\right>\mathrm{d}r
= \frac{\pi}{2}\frac{\phi(0)}{\int_0^{+\infty}\phi(k)\mathrm{d}k}
\end{equation}
where the wavenumber $k$ and the energy spectrum in wavenumber space $\phi(k)$ are defined as,
\begin{equation}
\left\{\begin{array}{l}
k = 2\pi f/\left<v\right> \\
\phi(k) = \frac{\left< v\right>}{2\pi}\phi(f)
\end{array}\right.
\end{equation}
For an ideal flat spectrum below $k_0$ and a $k^{-5/3}$ scaling above
$k_0$, we have,
\begin{equation}
\int_0^{+\infty}\phi(k)\mathrm{d}k = \frac{5}{2}\phi(0)k_0
\end{equation}
and therefore, one can derive the observed longitudinal integral scale
as $L_l = \frac{\left< v \right>}{10f_0}$ and then, assuming
homogeneous isotropic turbulence, the transverse integral scale $L_g$
as $L_g = L_l/2$. 

In our measurements, the low-frequency part of the spectrum is not
flat down to a few tens of mHz. Those small fluctuations only
represents some 0.1~\% of the mean velocity and therefore make little
change on the value of the turbulence intensity. They may come from small
and slow fluctuations of the forcing mean velocity rather than from
grid-generated turbulences.  Therefore, it is necessary to choose a
criterion to determine the corner frequency $f_0$.  We define it as
the frequency of the crossing of two power laws: one with a scaling
$f^{-5/3}$ fitted on the spectrum (inertial cascade) and one with an
arbitrary scaling $f^{-0.4}$ which roughly reproduces the resolved low
frequency part of the spectrum.
Values of corner frequencies and derived integral scales for each
spectrum are summarized in table \ref{tab:Ll}, including error estimates.
There was more noise during run 2, which explains the larger uncertainty on $f_0$.

\begin{table}
  \caption{\label{tab:Ll}Some integral scale measurements
    derived from the velocity power spectra obtained in run 1 (probe \ding{172})
    and run 2 (probe \ding{173}). For comparison, Comte-Bellot and Corrsin predictions gives $L_l = 2.6~\mathrm{mm}$ for run 1 and $L_l = 2.5~\mathrm{mm}$ for run 2.}
\begin{ruledtabular}
\begin{tabular}{cccccc}
Run & $x$ [mm] & $\left< v \right>$ [m/s] & $f_0$ [Hz] & $L_l$ [mm] & $L_g$ [mm] \\
\hline
\multicolumn{6}{c}{\emph{He~I \& He~II identical within error bars}} \\
1 & 540 & 3.3 & $140 \pm 25$ & $2.4 \pm 0.4$ & $1.2 \pm 0.2$ \\
1 & 540 & 2.5 & $105 \pm 25$ & $2.4 \pm 0.8$ & $1.2 \pm 0.4$ \\
1 & 540 & 1.7 & $74 \pm 25$ & $2.3 \pm 0.9$ & $1.15 \pm 0.45$ \\
\hline
\multicolumn{6}{c}{\emph{He~I only}} \\
2 & 470 & 4.2 & $154 \pm 50$ & $2.7 \pm 1.0$ & $1.3 \pm 0.5$ \\
2 & 470 & 2.5 & $98 \pm 40$ & $2.5 \pm 1.2$ & $1.25 \pm 0.6$
\end{tabular}
\end{ruledtabular}
\end{table}

\begin{figure}
\begin{center}
\includegraphics{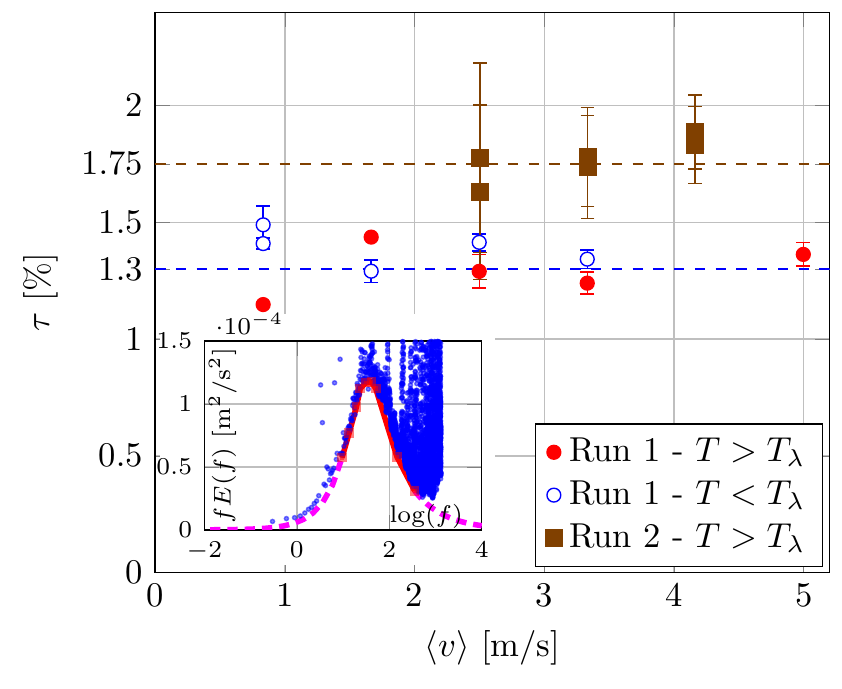}
\end{center}
\caption{Turbulence intensity $\tau$ measured for the two grid flow
  runs in the TSF wind tunnel, for various velocities and
  temperatures, computed using the integral of the energy
  spectrum. Inset: Estimation of the envelope of the energy
  spectrum. The area below the envelope is the energy of the velocity
  fluctuations.  The dots are experimental data points, the solid line
  is the estimated envelope below the spectrum and the dashed line is
  the extrapolated spectrum (flat spectrum in the low frequency limit
  and $f^{-5/3}$ scaling in the high frequency limit). The energy from
  the low frequency $f^{-0.4}$ increase is not taken in the turbulent
  energy estimate. However, this
  makes a relative difference smaller than a few percents in the final
  estimate.}\label{fig:TxTurb}
\end{figure}

To get the rms velocity fluctuations, or the turbulence
intensity, $\tau = \sqrt{\left< v'^2 \right>}/\left< v \right>$,
we calculate the area below $\phi(f)$ in a linear plot, or in
practice, the area below $f\cdot\phi(f)$ in a semilog plot, to have a
better estimate of the uncertainties (see inset of figure \ref{fig:TxTurb}).  
We also ignored the contribution of the low-frequency increase since
it is not expected to come from the turbulence cascade.

For run 1 ($x_1 = 540~\mathrm{mm}$), the measured turbulence intensity
is found to be $\tau_1 = (1.3 \pm 0.1)~\%$; for run 2 ($x_2 =
470~\mathrm{mm}$), $\tau_2 = (1.75 \pm 0.15)~\%$ (see figure
\ref{fig:TxTurb}).  The longitudinal integral scale are around $L_l =
2.5~\mathrm{mm}$ for both runs, the error bars make it impossible to
resolve the variation of $L_l$ between these two positions. As a first
result, we find that both quantities are consistent with Comte-Bellot
and Corrsin fit for classical grid flow. Besides, and more
importantly, we find that both the integral scale and the turbulence
intensity remain unchanged above and below the superfluid transition,
within relative experimental uncertainties of 8~\% for $\tau$ and 20~\%
for $L_l$.

From $\tau_1$ and $\tau_2$, we can estimate directly
the turbulence dissipation rate, $\epsilon$ from the turbulent kinetic
energy flux at position $x_1$ and $x_2$:
\begin{equation}
\epsilon \approx \left<v\right>^3\left|\frac{\partial\tau^2}{\partial x}\right|
\approx \left< v \right>^3\frac{(\tau_2^2-\tau_1^2)}{(x_2-x_1)}
\label{eq:epsilon}
\end{equation}
From the measured values, we can get $\partial\tau^2/\partial x \approx 0.0021~\mathrm{m^{-1}}$.
This is in good agreement, with less precise alternative estimation\cite{pope2000},
\begin{equation}
\epsilon \simeq 1.1 \frac{\left<v'^2\right>^{3/2}}{L_g} = 1.1\left< v\right>^3\frac{\tau^3}{L_g}
\end{equation}
where $1.1\tau^3/L_g$ lies in the range $0.0012 \mbox{ --- } 0.0045~\mathrm{m^{-1}}$ 

From $\epsilon$ and assuming isotropic and homogeneous turbulence, we
can compute the turbulence micro-scale $\lambda$ in He~I,
\begin{equation}
\epsilon = 15\nu\frac{\left<v'^2\right>}{\lambda^2}
\label{eq:lambda}
\end{equation}
The derived values of $\lambda$ lies in the range 70 ---
230~$\mathrm{\mu m}$ and $R_{\lambda} =
\lambda\sqrt{\left<v'^2\right>}/\nu$ in the range 60 --- 250.  We find
$R_{\lambda} \geq 100$ for most of our experimental conditions, which
is consistent with the assumption of developed grid turbulence above
the superfluid transition. Therefore, we expect the inertial range
energy spectrum to roughly follow the Kolmogorov prediction,
\begin{equation}
\phi(k) = C_k\epsilon^{2/3}k^{-5/3}
\end{equation}
On the inset of figure \ref{fig:spectra}, we plot the compensated
energy spectrum, 
\begin{equation}
\psi(k) = \epsilon^{-2/3}k^{5/3}\phi(k)
\end{equation}
From the value of
the ``plateau'', we can derive an estimate for the Kolmogorov constant,
$C_k$, in both He~I and He~II. We find values in the range $C_k = 0.3
\mbox{ --- } 0.4$. This is a one-dimensional Kolmogorov constant,
which can be related to the three-dimensional
Kolmogorov constant $C_{3d}$ assuming local isotropy,
\begin{equation}
C_{3d} = \frac{55}{18}C_k
\end{equation}
We find that the three-dimensional Kolmogorov constant lies in the
range $0.9 \mbox{ --- } 1.2$. 

Previous normal fluid grid flow
experiments\cite{comteBellot1971,gadElHak1974,gibson1963,schedvin1974}
have reported measured values of the Kolmogorov constant
scattered\cite{sreenivasan1995} around $C_{3d} = 1.5$, in the window
$C_{3d} = 1.0 \mbox{ --- } 1.74$ (ie. $C_k = 0.33 \mbox{ --- } 0.57$).
The value that we find is close to the smaller values reported
in the literature. Our emphasis will not be on the actual value that
we have measured. Indeed, the latter can be affected by systematic
errors, such as systematic bias on the probe calibration. However,
it is quite remarkable that our measure of the Kolmogorov constant in
He~II down to 2.0~K coincides with the value measured in He~I within
30~\% relative error margin.

\section{High turbulence intensity flows}

We report two sets of high turbulence intensity flows:
measurements done in the TSF wind tunnel in the near wake of
a cylinder (see schematics of run 1 on figure \ref{fig:schemaTSF}-a) and
measurements done in the NÉEL wind tunnel, sketched on figure
\ref{fig:schemaTSF}-c and described in more details elsewhere\cite{roche2007}.
The main advantage of such flows is a better signal-to-noise ratio. However, the
turbulence is less homogeneous and less isotropic, especially in the near wake flow.

\subsection{Near wake flow}

\begin{figure}
\begin{center}
\hspace*{\stretch{1}}\includegraphics[height=3cm]{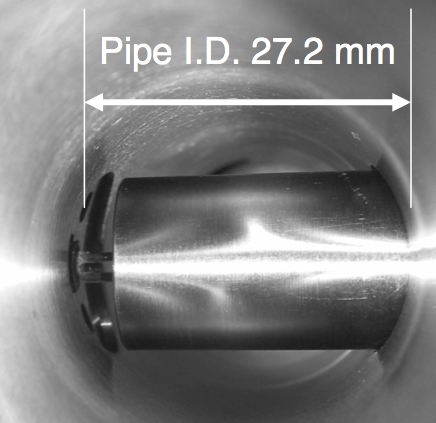} \hfill
\includegraphics[height=3cm]{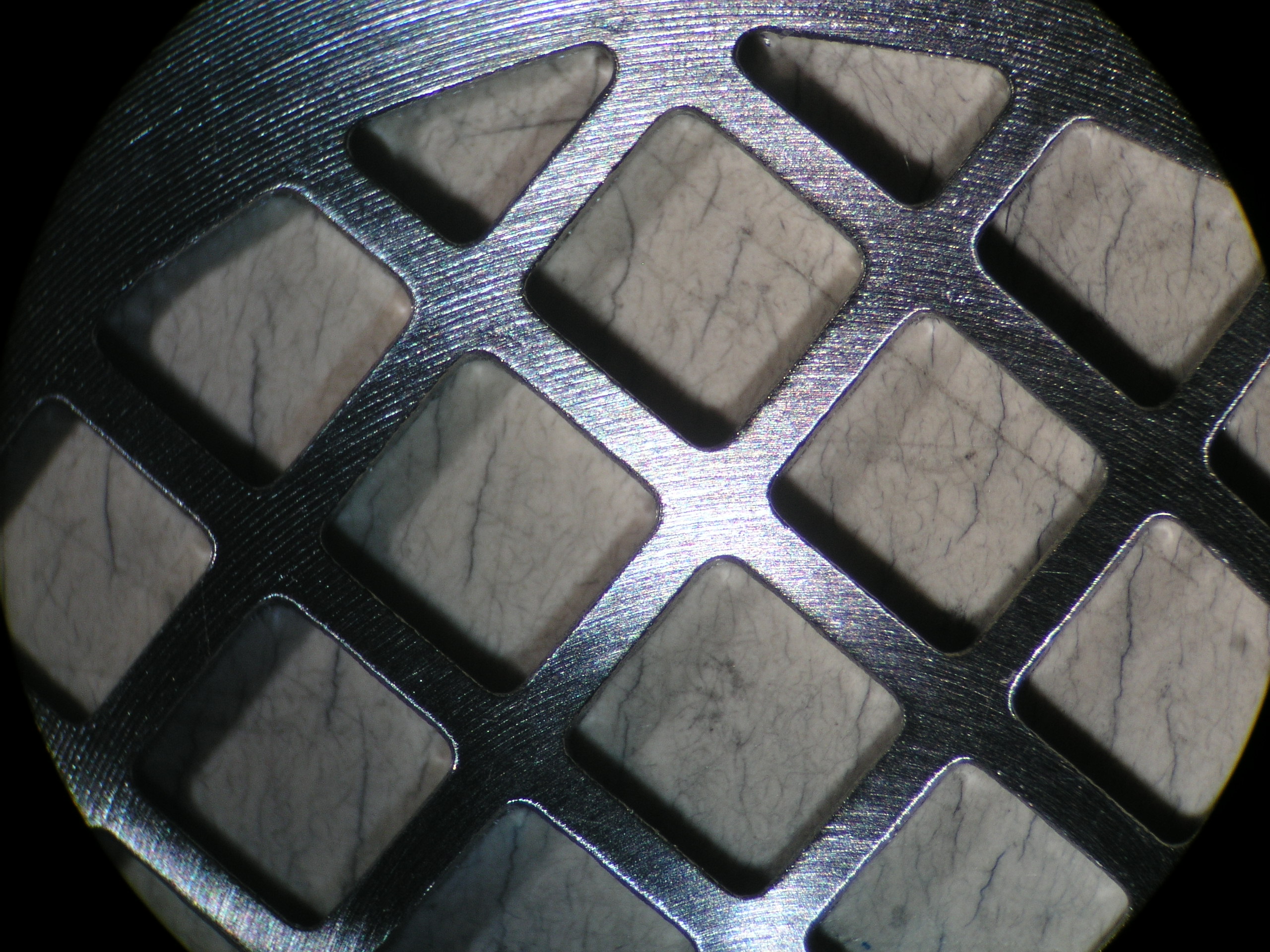}\hspace*{\stretch{1}}
\end{center}
\caption{Left: picture of the removable cylinder in the TSF wind tunnel. The angle between
the probe and the axis of the pipe is 17°. Right: picture of the grid.}\label{fig:cylindre}
\end{figure}

The cylinder used in the TSF wind tunnel was originally designed to
protect a hot-wire during the filling of the cryogenic loop, in
particular to avoid droplets from colliding with the wire.  Therefore,
the dimensions are not designed to produce fully developed wake
turbulence. As shown on table \ref{table:dimTSF}, the wake cylinder
diameter $\varnothing_c$ is 15.3~mm for a pipe diameter
$\varnothing_p$ of 27.2~mm, leading to a significant wall confinement.
Besides, the cylinder length is slightly smaller than the pipe
diameter as shown on figure \ref{fig:cylindre}. The dimentionless
distance between the cylinder axis and the sensor, $L_1/\varnothing_c$
is $4.0 \pm 0.3$. The cylinder Reynolds number $\varnothing_c \left< v
\right>/\nu$ falls in the range $3\times 10^5$ --- $2\times 10^6$,
where $\left< v \right>$ is estimated upstream (or downstream) from
the cylinder, and not on the constriction where $\left< v \right>$ is
larger.  In a less confined geometry, the Strouhal number,
\begin{equation}
\mathrm{St} = \frac{f_v \varnothing_c}{\left< v \right>}
\end{equation}
where $f_v$ is the frequency of vortex shedding, is undefined at such Re in classical
fluids\cite{lienhard1966}. Finally, we point that this flow geometry
can lead to large angle of attack on the probe.

\begin{figure}
\begin{center}
\includegraphics{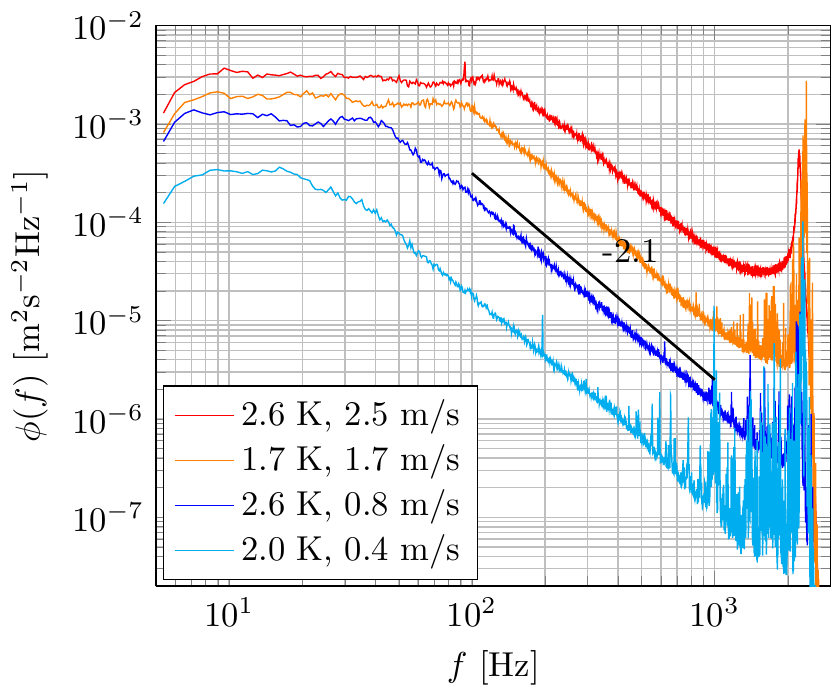}
\end{center}
\caption{Velocity spectra in the near wake of a cylinder in the TSF
  wind tunnel both above and below the superfluid transition with mean
  velocity increasing from bottom to top. The high frequency peak near
  2~kHz is the sensor Helmholtz frequency.}\label{fig:sillages}
\end{figure}

\begin{figure}
\centering\includegraphics{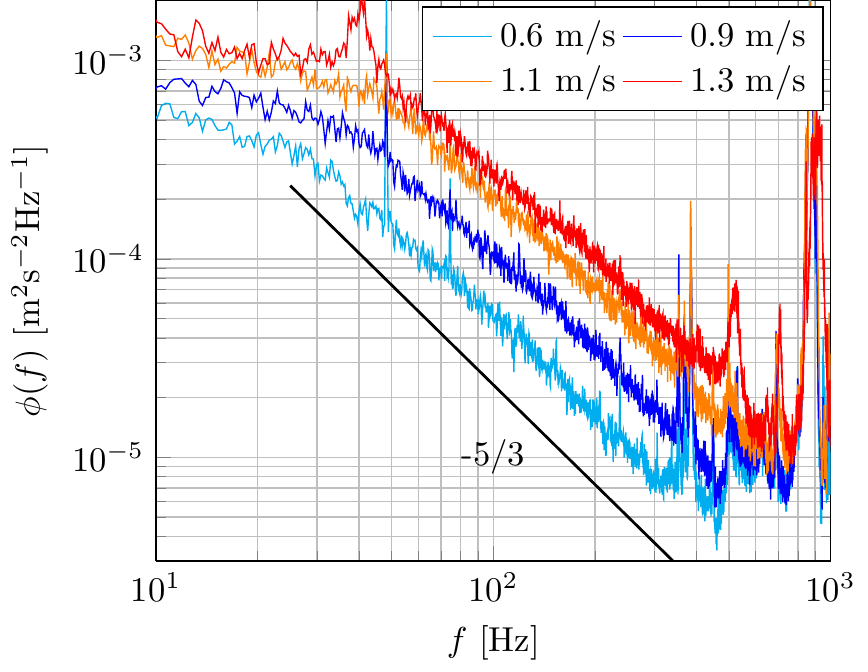}
\caption{Velocity spectra for $T=1.55~\mathrm{K}$ in the NÉEL wind
  tunnel for four mean velocities below the superfluid transition.
  The low frequency corners give an estimate of the longitudinal integral scale
  $L_g = 2~\mathrm{cm}$.}\label{fig:spectreNEEL}
\end{figure}

\begin{figure}
\centering\includegraphics{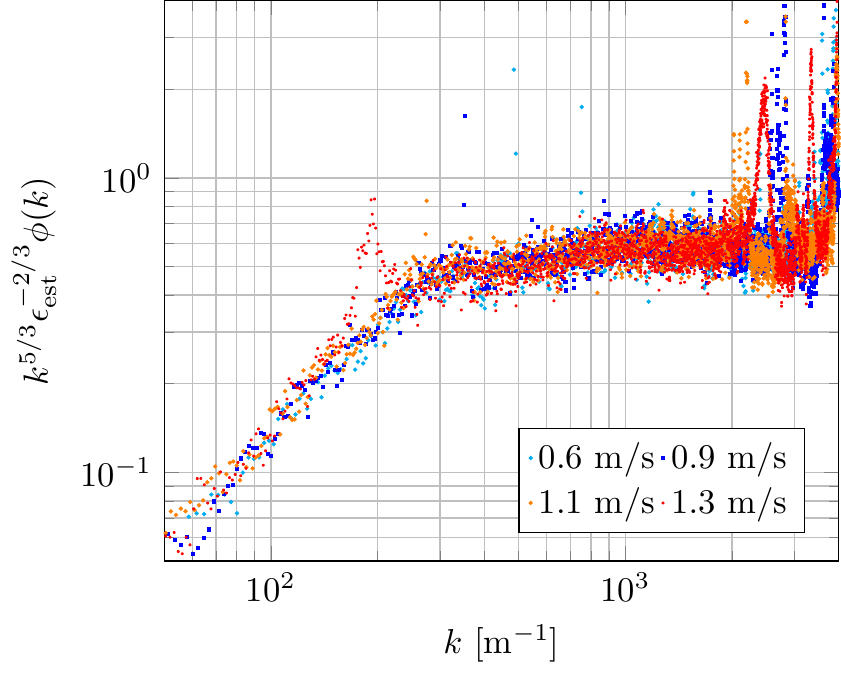}
\caption{Same data as figure \ref{fig:spectreNEEL} plotted in
a compensated fashion (see text).}\label{fig:spectreCompNEEL}
\end{figure}

Figure \ref{fig:sillages} shows spectra in the near wake of the
cylinder in both He~I and He~II. No sharp Strouhal peak is visible,
either above or below the superfluid transition.  The slope is steeper
than -5/3. One possible explanation is that the
spectral distribution of energy right after the obstacle is concentrated
at the largest scales and by the time the probe is reached, it has not developped
yet into the Kolmogorov cascade.
As another possible explanation, we also point out that velocity spectra in
strongly inhomogeneous classical flows, in particular near a stable
vortex, are known\cite{simand2000} to scale like $f^{-\alpha}$, with
$\alpha$ in the range 1.65 --- 2.50.  In any case, our result shows
that the indistinguishability between He~I and He~II does not require
an equilibrium state in the sense of Kolmogorov.

\subsection{Chunk turbulence}

The NÉEL wind tunnel is placed in a saturated liquid helium bath (see
figure \ref{fig:schemaTSF}-c). The temperature is controlled mainly by
the bath pressure and fine-tuned by a temperature regulator. The data
discussed here are obtained at $T=1.55~\mathrm{K}$, which corresponds
to a superfluid fraction $\rho_s/\rho \approx 86~\%$. Above the
superfluid transition, bubbles are likely to appear in saturated
baths. Therefore we only report measurements below the superfluid
transition, where the absence of thermal gradients prevents the
forming of bubbles.  The turbulence is generated by a continuously
powered centrifugal pump and probed by stagnation pressure probe
\ding{175} and a local quantum vortex lines density probes in a
23~mm-diameter, 250~mm-long brass pipe, located upstream from the
pump. The analysis of the quantum vortex lines density results are
discussed in a previous paper\cite{roche2007,roche2008}.  The useful
range of velocity is 0.25 --- 1.3~m/s. The typical turbulence
intensity is roughly constant in this range of parameters.  Its value
is $(18 \pm 1)~\%$ if we choose to remove the energy that comes from
the low-frequency variation of the mean velocity, like we did in the
previous parts; or in the range 25~\% --- 35~\%, if we choose to keep
all the measured energy, like was done in the previous
paper\cite{roche2007}. The superfluid Reynolds number
$\mathrm{Re}_{\kappa} = \varnothing V/\kappa$ falls in the range
$6\times 10^4$ --- $3\times 10^5$.

Figure \ref{fig:spectreNEEL} shows spectra obtained in the NÉEL wind
tunnel in He~II.  They show one decade of $f^{-\alpha}$
scaling, with $\alpha = 1.55 \mbox{ --- } 1.69$. This is compatible
with a Kolmogorov -5/3 turbulent cascade with a relative experimental
error bar of less than 7~\% on the exponent. The compensated spectra are shown on
figure \ref{fig:spectreCompNEEL} using $\epsilon_{\text{est}} =
\left<v'^2\right>^{3/2}/L_g$ and $L_g = 1~\mathrm{cm}$. 
From the value of the ``plateau'', we find a one-dimensional
Kolmogorov constant around $0.5$. Although the ``chunky'' aspect of
the flow prevents to speculate on its value, we note that it is in
good agreement with values in a classical flows.

\section{Conclusions \& perspectives}

We have done systematic superfluid velocity measurements in three
different highly turbulent flows.  The upper inertial range of the
turbulent cascade was resolved with various anemometers based on
stagnation pressure probes.  We found that the second order statistics
of the superfluid velocity fluctuations does not seem to differ
from those of classical turbulence down to the precision of our measurements.

It is worth pointing that non conventional velocity statistics have
been recently reported in superfluid flows, both in an
experimental\cite{paoletti2008} and a numerical\cite{white2010}
study. These studies where conducted at a much lower effective
Reynolds number and the probing of the flow velocity was done at a
scale where quantum effects are prevalent. In the present work, the
characteristic length scale of quantum effect is much lower than the
probe resolution. For example, in the NÉEL flow, the typical distance
between two neighbouring quantum vorticies is a few microns\cite{roche2008}, 
to be compared with the probe resolution of 1~mm typically.

To go further into the physics of quantum turbulence, it would be
necessary to resolve the small scales of a high-Reynolds number
flow. To do this at given Reynolds number, one should either increase
the cut-off scale by scaling up the experiment or decrease the size of
the probe.  However, it is delicate to reduce the size of stagnation
pressure probes below-say-$200~\mathrm{\mu m}$.  One alternative is to
design new types of probes --- for example, adapting cantilever-based
anemometers\cite{barth2005} to low temperatures.

\begin{acknowledgments}
  This work would not have been possible without the precious help and
  support of M. Bon Mardion, A. Forgeas, P. Roussel and J.-M. Poncet
  (SBT) and G. Garde, A. Girardin, C. Guttin and Ph. Gandit (Institut
  Néel) nor without the financial support of the ANR (grant
  ANR-05-BLAN-0316, ``TSF'') and the Région Rhône-Alpes.

  We thank also R. Kaiser for his contribution during data
  acquisition, T. Haruyama (KEK, Japan) for support with several
  pressure transducers and L. Chevillard and F. Chillà (ENS Lyon) for
  fruitful discussions.

\end{acknowledgments}

\appendix
\section{Derivation of the stagnation pressure signal}
\label{app:Calculs}

We consider the total pressure $U(t)$ measured by a stagnation pressure
probe in a classical incompressible fluid,
\begin{equation}
U(t) = \frac{1}{2}\rho v(t)^2 + P(t)
\label{eq:signal}
\end{equation}
where $\rho$ is the density of the fluid, $v(t)$ the local velocity
and $P(t)$ the local static pressure. Equation \ref{eq:signal} can
be rewritten using Reynolds decomposition $v(t) = \left< v \right> +
v'(t)$ and $P(t) = \left<P\right> + P'(t)$,
\begin{equation}
U(t) = \frac{1}{2}\rho \left< v \right>^2 + \left<P\right> + \rho\left< v\right> v'(t) + P'(t) + \frac{1}{2}\rho v'(t)^2
\label{eq:allTerms}
\end{equation}
We recall the definition of the turbulence intensity $\tau$,
\begin{equation}
\tau = \frac{\sqrt{\left< v'^2 \right>}}{\left< v \right>}
\end{equation}

The typical magnitude of the static pressure fluctuation $P'$ can be estimated
for isotropic and homogeneous turbulence\cite{uberoi1953,batchelor1951,schumann1978},
\begin{equation}
\frac{\sqrt{\left< P'^2 \right>}}{\frac{1}{2}\rho\left< v'^2 \right>} \approx 1.4
\label{eq:rmsP}
\end{equation}
Therefore, the terms of equation \ref{eq:allTerms} can be divided in orders of $\tau$,
\begin{equation}
\left\{ \begin{array}{lclcl}
U_0 & = & \frac{1}{2}\rho\left< v \right>^2 + \left<P\right> & = & \mathcal{O}(1) \\
U_1(t) & = & \rho\left< v \right>v'(t) & = & \mathcal{O}(\tau) \\
U_2(t) & = & P'(t) + \frac{1}{2}\rho v'(t)^2 & = & \mathcal{O}(\tau^2)
\end{array}\right.
\end{equation}
$U_0$ is a constant offset, used only for calibrating the probe, $U_1(t)$ is the signal
of interest and $U_2(t)$ is the second order corrective term, considered as a spurious
signal for stagnation pressure probes.
The relative weight of $U_2$ versus $U_1$ can be estimated versus the turbulent
intensity $\tau$,
\begin{equation}
\begin{array}{rcc}
U_1(t) & = & \hspace{-0.6cm}\rho\left< v \right> v'(t) \sim  \rho \left< v \right>^2 \tau \\
U_2(t) & = & \left\{ \begin{array}{rcl}
    P'(t) & \sim & 0.7 \rho\left< v \right>^2 \tau^2 \\
    \frac{1}{2}\rho v'(t)^2 & \sim & 0.5 \rho \left< v \right>^2 \tau^2
    \end{array}\right.
\end{array}
\label{eq:weight}
\end{equation}
Some values are given in table \ref{tab:terms}.  We can see that for
turbulence intensity larger than 20~\%, like those obtained in Von
Kármán cells, and in wake or ``chunk'' flows, almost 30~\% of the
measured signal comes from second order correction terms. However, for
turbulence intensities of grid flows, less than 2~\% in our case, more
than 96~\% of the measured signal comes from the linear velocity term.

\begin{table}
\caption{\label{tab:terms}Typical relative weight of each term contributing to the
signal measured by a stagnation pressure probe
for various turbulence intensity $\tau$ using the estimate (\ref{eq:weight})}
\begin{ruledtabular}
\begin{tabular}{cccc}
$\tau$ & $\rho\left<v\right>v'(t)$ & $P'(t)$ & $\frac{1}{2}\rho v'(t)^2$ \\
\hline
1~\% & 98.8~\% & 0.7~\% & 0.5~\% \\
2~\% & 97.6~\% & 1.4~\% & 1~\% \\
10~\% & 89.2~\% & 6.3~\% & 4.5~\% \\
20~\% & 80.6~\% & 11.3~\% & 8.1~\% \\
30~\% & 73.5~\% & 15.5~\% & 11.0~\%
\end{tabular}
\end{ruledtabular}
\end{table}

\bigskip

\noindent \textbf{\textsf{REFERENCES}}\par


%

\end{document}